%% file: main.tex
\documentclass[conference]{IEEEtran}
\IEEEoverridecommandlockouts
\usepackage{cite}
\usepackage{amsmath,amssymb,amsfonts}
\usepackage{algorithm}
\usepackage{graphicx}
\usepackage{textcomp}
\usepackage{tikz}
\usepackage{pgfplots}
\usepackage{xcolor}
\usepackage{mathtools}
\usepackage{algpseudocode}

\usepackage[acronyms,nonumberlist,nopostdot,nomain,nogroupskip]{glossaries}
\input{acronyms.tex}
\def\BibTeX{{\rm B\kern-.05em{\sc i\kern-.025em b}\kern-.08em
    T\kern-.1667em\lower.7ex\hbox{E}\kern-.125emX}}

\makeatletter
\newcommand\setalgorithmcaptionfont[1]{%
  \let\my@floatc@ruled\floatc@ruled          
  \def\floatc@ruled{%
    \global\let\floatc@ruled\my@floatc@ruled 
    #1\floatc@ruled}}
\makeatother

\begin{document}

\title{A New Scheduler for URLLC in 5G NR IIoT Networks with Spatio-Temporal Traffic Correlations
}




\author{\IEEEauthorblockN{ Sara Cavallero$^{\circ }$, Nicol Sarcone Grande$^{\circ }$, Francesco Pase$^{\star }$\\
Marco Giordani$^{\star }$, Joseph Eichinger$^{\dagger }$, Roberto Verdone$^{\circ }$, Michele Zorzi$^{\star }$\medskip} \IEEEauthorblockA{
 $^{\circ}$WiLab/CNIT and University of Bologna, Italy. \\
 Email: \texttt{\{s.cavallero; nicol.sarconegrande; roberto.verdone\}@unibo.it}\\
 $^{\star}$WiLab/CNIT and University of Padova, Italy. \\Email: \texttt{\{pasefrance; giordani; zorzi\}@dei.unipd.it}\\
 $^{\dagger}$Huawei Technologies, Munich Research Center, Germany. \\Email: \texttt{joseph.eichinger@huawei.com}}
 }
\maketitle

\begin{abstract}
This paper explores the issue of enabling \gls{urllc} in view of the spatio-temporal correlations that characterize real \gls{5g} \gls{iiot} networks. 
In this context, we consider a common Standalone Non-Public Network (SNPN) architecture as promoted by the 5G Alliance for Connected Industries and Automation (5G-ACIA), and propose a new variant of the 5G NR \gls{sps} to deal with uplink traffic correlations. 
A benchmark solution with a ``{smart}'' scheduler (SSPS) is compared with a more realistic adaptive approach (ASPS) that requires the scheduler to estimate some unknown network parameters.
We demonstrate via simulations that the 1-ms latency requirement for URLLC is fulfilled in both solutions, at the expense of some complexity introduced in the management of the traffic. 
Finally, we provide numerical guidelines to dimension IIoT networks as a function of the use case, the number of machines in the factory, and considering both periodic and aperiodic traffic.
\end{abstract}
			\begin{tikzpicture}[remember picture,overlay]
	\node[anchor=north,yshift=-10pt] at (current page.north) {\parbox{\dimexpr\textwidth-\fboxsep-\fboxrule\relax}{
			\centering\footnotesize This paper has been accepted for presentation at the 2023 IEEE International Conference on Communications (ICC). \textcopyright 2023 IEEE. \\
			Please cite it as: S. Cavallero, N. Sarcone Grande, F. Pase, M. Giordani, J. Eichinger, R. Verdone, M. Zorzi, “A New Scheduler for URLLC in 5G NR IIoT Networks with Spatio-Temporal Traffic Correlations,” IEEE International Conference on Communications (ICC), Rome, Italy, 2023.\\
	}};
\end{tikzpicture}
\glsresetall


\section{Introduction}
The fourth Industrial Revolution (Industry 4.0) is driven by several \gls{5g} technologies such as \gls{ai} and robotics, and aims at improving the efficiency, security, and revenue of \gls{iiot} applications \cite{ICTSurvey, Industry_4.0}.  
For some \gls{iiot} processes, \gls{urllc} is required, i.e., network reliability up to 99.99999\%, and \gls{e2e} latency below 1 ms~\cite{ 9057670, 5gaciaiiot,boban2021predictive}. 

From the network architecture point of view, the \gls{3gpp} introduces in Rel-16 (and future specifications) the paradigms of \gls{snpn} and \gls{pninpn} to promote \gls{urllc} in \gls{iiot} scenarios~\cite{3gpp.23.501,snpn}. The \gls{3gpp} \gls{snpn} paradigm has also been explored by the \gls{5gacia} to recommend promising network architectures able to satisfy \gls{iiot} application requirements~\cite{5gaciaarchitecture}.
With regard to resource allocation, the \gls{3gpp} NR supports \gls{sps} in the \gls{ul}~\cite{3gpp.38.321}, in which the network pre-allocates radio resources, avoiding the need for the \glspl{ue} to receive uplink grants before transmission as in conventional {grant-based} scheduling. 
Many works have analyzed and demonstrated the potential of \gls{sps} to reduce the \gls{ul} latency~\cite{feng2019predictive, arnjad2018latency, shahsin2020adaptive}, also against some other contention-based solutions which prioritize latency at the expense of reliability~\cite{jacobsen2017system, lucas2019capacity}. 
However, most of the prior art neglects the impact of the \gls{5g} protocol stack, thus considering the air latency instead of the more representative \gls{e2e}~latency, or introduces some assumptions in the traffic model. 
For~example, in our previous contribution~\cite{eucnc}, we assumed that \glspl{ue} in different machines generate packets simultaneously and according to pre-defined periodicity and size, thus neglecting the complexity of \gls{iiot} scenarios where different types of traffic may coexist.



To fill these gaps, in this paper we introduce a new, but more realistic, traffic model in which industrial machines and users activate and generate traffic, respectively, based on some specific temporal and spatial correlations.
Based on that, we propose a new implementation of the \gls{sps}  for \gls{iiot} networks based on a \gls{5gacia} \gls{snpn} architecture, able to predict and exploit the temporal/spatial correlations of the traffic, and to pre-allocate resources accordingly.
We consider two schemes: (1) the \gls{ssps} (our benchmark), in which a ``{smart}'' \gls{gnb} is assumed to have full knowledge of the system model; and (2) the \gls{asps} in which the \gls{gnb} implements a heuristic algorithm to estimate the system parameters for resource allocation.
The performance of the proposed schemes is studied in terms of \gls{e2e} latency against some baselines, as a function of the type of traffic (periodic or aperiodic), the offered traffic, the number of users, the bandwidth, and considering two IIoT use cases with different requirements.
To do so, we develop a custom full-stack simulator implementing the \gls{3gpp} channel model proposed in \cite{3gpp.38.901}, already 
tailored to IIoT environments~\cite{eucnc}, and model the total latency taking into account the time for the protocol and the control operations at the transmitter, data transmission and propagation, \gls{rf} processing at the receiver, and the delay introduced by the core network. 


The rest of the paper is organized as follows. In Sec.~\ref{sec:e2e_system_model} we describe our system model and simulator, in Sec.~\ref{sec:SPS_scheduling} we present our proposed \gls{sps} schemes for \gls{iiot} networks considering traffic correlations, in Sec.~\ref{sec:numerical_results} we introduce our system parameters and discuss the numerical results, and in Sec.~\ref{sec:conclusions} we conclude with suggestions for future research.
\begin{figure}[!t]
\begin{center}
	\includegraphics[width=0.8\columnwidth	]{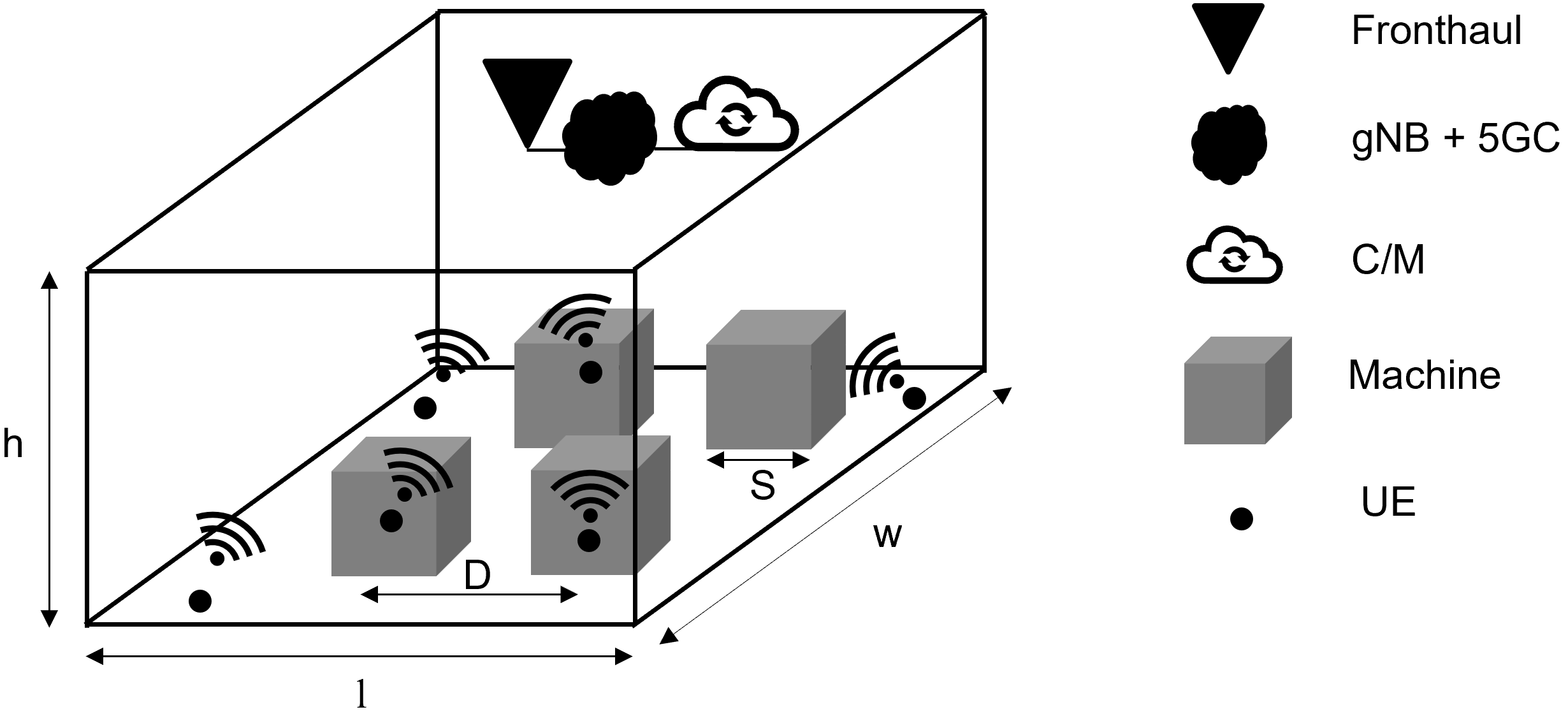}
	\caption{The baseline 5G-ACIA network architecture~\cite{5gaciaarchitecture}.}
	\label{fig:baseline_network_architecture}
\end{center}
\vspace{-3ex}
\end{figure}

\section{End-to-End System Model}
\label{sec:e2e_system_model}
\subsection{Network Architecture}
The reference scenario is a factory with ``{machines}'' (industrial assets in charge of different tasks), connected to an \gls{snpn}, which is a \gls{5g} remote and private network with a reserved \gls{ran} and \gls{5gc}. 
Our baseline \gls{snpn} architecture is designed according to the \gls{5gacia} recommendation, as illustrated in Fig.~\ref{fig:baseline_network_architecture}, where the \gls{cm} is the \gls{iiot} central entity responsible for monitoring and controlling the machines from the \gls{5gc}.

\subsection{Deployment Model}
The \gls{gnb} and its fronthaul are located at the center of the ceiling of the factory, as shown in Fig.~\ref{fig:baseline_network_architecture}. The factory floor is modeled as a parallelepiped of length $l$, width $w$, and height $h$, as indicated in~\cite{5gaciarequirement}. 
We consider $M$ machines, modeled as fixed cubes of size $S$, and deployed on the factory floor according to a uniform distribution ensuring a given inter-machine distance $D$ (evaluated from halfway across the lower base) and a minimum number of machines $M_{\rm min}$ (where both $D$ and $M_{\rm min}$ are input parameters of our simulator). 
Likewise, $N$ \glspl{ue} are distributed following the same method, at a maximum height $S$, and can be placed onboard industrial machines, on top of walls, on metal slabs, or on robots.
Inside the factory, several ''obstacles'' may obstruct the communication between the \glspl{ue} and the \gls{gnb}. 

\subsection{Channel Model}
The channel is characterized  based on the \gls{3gpp} \gls{inf} model for \gls{iiot} networks~\cite[Table 7.2-4]{3gpp.38.901} and is implemented as in our previous work ~\cite{eucnc}. Specifically, the model suggests four \gls{inf} scenarios, depending on the density of machines and the height of the \glspl{ue} and of the \gls{gnb} with respect to the ground. Then, the path loss depends on the condition of the channel (\gls{los} or \gls{nlos}), while the quality of the received signal is assessed in terms of the \gls{snr}.

\subsection{Traffic Model}
\label{sub:correlated_traffic}
Unlike in previous work \cite{eucnc}, the traffic is generated according to pre-defined IIoT-specific correlations. 
As such, the $M$ machines are organized in $n_{\rm lines}$ production lines, and each \gls{ue} is associated with the nearest machine. 
When a machine in a production line activates, its \glspl{ue} produce data traffic for an entire {activation period} of duration $\tau_{\rm on}$. 
The traffic model accounts for both temporal and spatial correlations. 
In principle, we investigate two different types of correlation:
\begin{itemize}
    \item Inter-machine correlation, which refers to the way machines activate. When an event occurs, one machine per production line is activated. Then, machines in the same production line activate in sequence.
    \item Intra-machine correlation, which refers to the way \glspl{ue} associated with each active machine generate data. After one machine per line is activated, the \glspl{ue} associated with those machines activate too, each with probability~\textit{p}. Once active, each \gls{ue} generates a flow of packets according to some statistics (e.g., in terms of the inter-packet interval and/or the packet size) for a whole $\tau_{\rm on}$. We consider periodic traffic, generated at constant periodicity $\tau$, or aperiodic traffic, in which the inter-data-burst interval changes according to a uniform random variable within the interval [$t_{\rm min}$, $t_{\rm max}$]. 
\end{itemize}


\section{5G-NR Uplink Scheduler \\ with traffic correlations} 
\label{sec:SPS_scheduling}
The \gls{e2e} latency is affected by the implementation of the 5G \gls{us}, located at the \gls{gnb}.
After a brief overview on the baseline 5G NR SPS (Sec.~\ref{sec:overview}), in Sec.~\ref{sec:semi-persistent_scheduling} we present our new SPS designs for correlated traffic.


\subsection{5G NR SPS General Overview}
\label{sec:overview}

Four messages are exchanged for the 5G NR scheduling: (1) each \gls{ue} uses the \gls{pucch} to ask the \gls{us} for being scheduled; (2) the \gls{gnb} communicates via the \gls{pdcch} to the \glspl{ue} which resources can be used for transmission; (3) the \glspl{ue} transmit their data blocks through the \gls{pusch}; (4) the \gls{gnb} provides the communication acknowledgment via \gls{harq}.
In the frequency domain, the overall bandwidth $B$ is split into $n_{\rm RB}$ \glspl{rb}, 
each made of 12 \gls{ofdm} subcarriers~\cite{5gbook2020}, where \gls{pucch}, \gls{pdcch} and \gls{harq} signals occupy one \gls{rb}, whereas the number of \glspl{rb} in the \gls{pusch} depends on the modulation order, the number of mapped data blocks, and their size. 
On the contrary, the time domain is organized in \gls{ofdm} symbols, where 7 \gls{ofdm} symbols ($n_{\rm os}$) define a \gls{su}: \glspl{pucch} and \glspl{pdcch} occupy one \gls{su} each, while \glspl{pusch} and \glspl{harq} form one \gls{su}, where one \gls{ofdm} symbol is spent to switch from transmission to reception and vice versa under the assumption of half-duplex communication.

Every $T_{\rm IP}$, i.e., the inter-PUCCH time, each \gls{ue} asks for resources by means of the \gls{pucch}, after which the \gls{us} employs one \gls{su} to process the request and allocate resources accordingly.
With \gls{sps}, the US (1) estimates the times at which new data blocks will be generated until the next \gls{pucch} opportunity; (2) allocates time/frequency resources based on the SU in which they will be served (where data blocks generated at $SU_{i}$ will be assigned to $SU_{ i+1}$); and (3) notifies the \glspl{ue} via the \gls{pdcch}, which takes another \gls{su}.
\begin{figure}[!t]
\begin{center}
	\includegraphics[width=0.9\columnwidth]{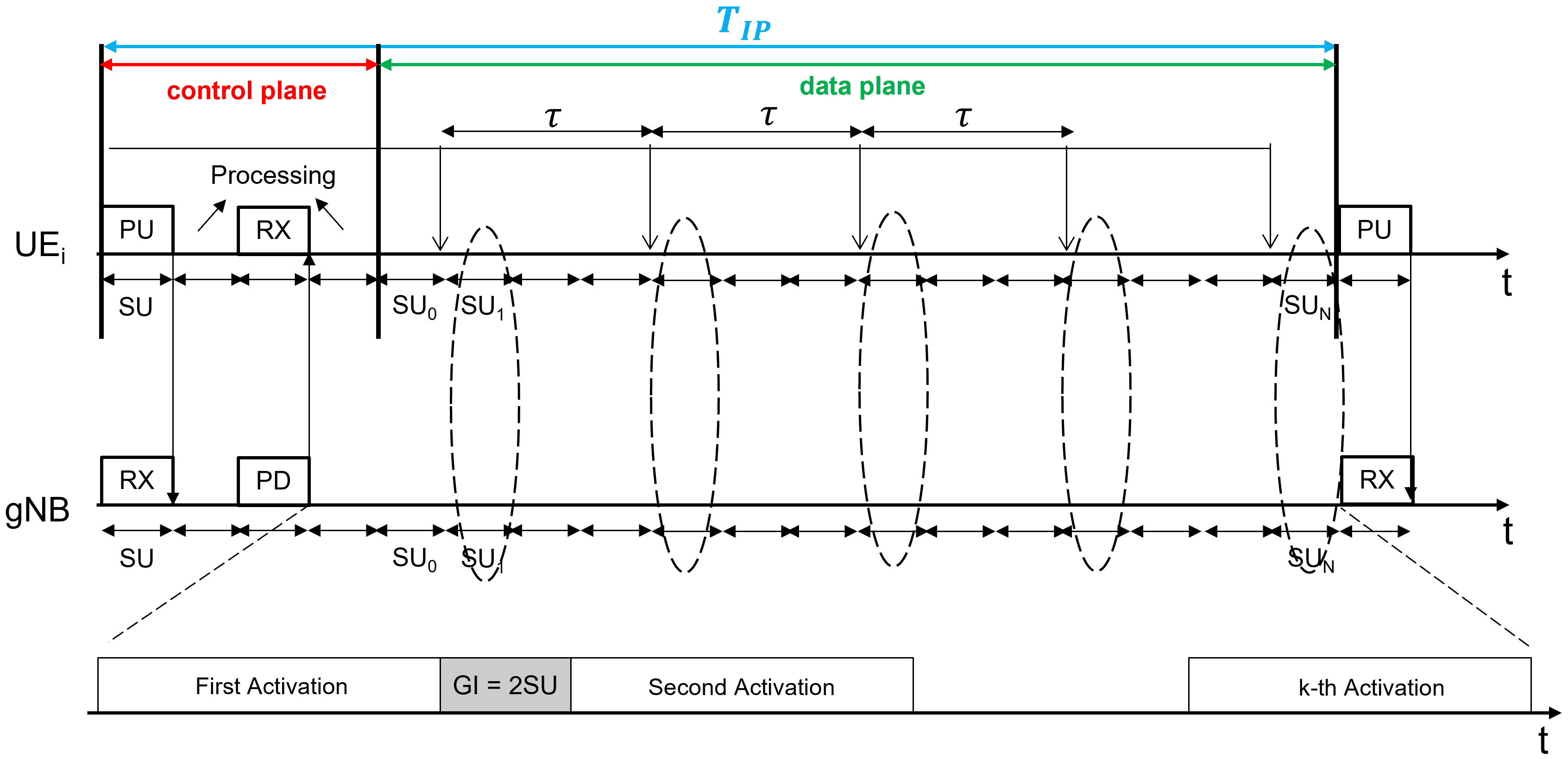}
    \caption{Timing diagram of the proposed SPS scheme for correlated traffic. The \gls{pusch} opportunities are allocated to the $i$-th \gls{ue} immediately after a new data block generation.}
	\label{fig:SPS}
\end{center}
\vspace{-3ex}
\end{figure}
\newline
In our simulator, the \gls{us} assigns resources according to the following criteria, in sequential~order:
\begin{enumerate}
    \item \gls{edf}, i.e., \glspl{ue}' requests that~are closer to the 1-ms latency~requirement are prioritized.
    \item \gls{ff}, i.e., \glspl{ue} are served up to a given minimum level (the bucket size $B_s$), to promote fairness during transmission. This approach avoids that some \glspl{ue} monopolize the channel at the expense of the others. If there are still resources available, the \glspl{ue}' requests beyond the bucket size are processed.
\end{enumerate}

Notice that, 
when considering correlated traffic, at each \gls{pucch} opportunity only a fraction of the machines in the factory is active, so only the \glspl{ue} associated with those specific machines can send the \gls{pucch}. 
Ideally, if the \gls{us} knows the spatio-temporal statistics of the correlated traffic, it can schedule resources even without explicit scheduling requests from the \glspl{ue} via the \gls{pucch}, as we will propose to do in the following section.

\subsection{Proposed SPS for IIoT Correlated Traffic}
\label{sec:semi-persistent_scheduling}
In~\cite{eucnc} we showed that \gls{sps} may be inefficient when the  traffic  is  aperiodic, due to estimation errors in the \gls{us}.
In order to deal with traffic correlations, we introduce the following assumptions for aperiodic \glspl{ue}.
\begin{itemize}
    \item To avoid packet accumulation in the queue, and the possibility that packets are not transmitted on time before the \gls{ue} is switched off, each UE can generate at most two packets during $\tau_{\rm on}$, as follows: 
\begin{itemize}
    \item The first packet is generated randomly within the interval $\left[t_{\rm min},\, ({t_{\rm min}+t_{\rm max}})/{2}\right]$.
    \item The second packet is generated randomly within the interval [$t_{\rm min}$, $t_{\rm max}$], provided that it is within the \gls{ue}'s activation period. 
\end{itemize}
\item The \gls{us} schedules three \glspl{su} per activation period for each aperiodic UE: one \gls{su} at time $({t_{\rm min}+t_{\rm max}})/{2}$, and two \glspl{su} in the second-last and last resources of the current activation period, respectively. 
\item If many \glspl{ue} have to send packets in the last two assigned \glspl{su}, more resources 
may be needed to complete their transmissions, which may involve using some of those already allocated in the following $\tau_{\rm on}$.
Therefore, a {\gls{gi}} of two \glspl{su} has been introduced in the data plane at the end of each activation period, to complete ongoing transmissions. The \gls{gi} ensures that \glspl{ue}' transmissions in consecutive activation periods do not overlap, as depicted in Fig.~\ref{fig:SPS}.
The effect of the \gls{gi} affects the definition of $T_{\rm IP}$. In particular, we have that 
\begin{equation}
\label{eq:t_ip}
    T_{\rm IP} =  (\tau_{\rm on} + 2\cdot \text{SU}) \cdot n_{\rm on},
\end{equation}
where $n_{\rm on}$ is the number of activations during $T_{\rm IP}$, and is  
an input parameter which allows some machines to re-activate within the same $T_{\rm IP}$, especially when $M$ is~small.
\end{itemize}

Based on the above assumptions, we propose two implementations of the \gls{sps} for correlated traffic.
\smallskip

\subsubsection{\Acrfull{ssps}}
\label{sub:semi-persistent gNB smart}

\noindent The ``smart \gls{gnb}'' is fully aware of the \gls{e2e} system model, including:
\begin{itemize}
    \item the number of production lines ($n_{\rm lines}$) inside the factory;
    \item the number of machines ($M$) in each production line;
    \item both the number and the ID of the \glspl{ue} associated with each machine;
    \item the type of traffic generated by the \glspl{ue} (i.e., periodic or aperiodic), and the data block size;
    \item the activation period ($\tau_{\rm on}$) of the machines;
    \item the number of activations per $T_{\rm IP}$ ($n_{\rm on}$).
\end{itemize}
When the \gls{gnb} receives the \gls{pucch} requests from the active \glspl{ue}, 
it can immediately identify the active machines, and those that will activate in the following $\tau_{\rm on}$.
As such, the \gls{gnb} can accurately predict the entire flow of activations, and schedule resources accordingly. 
\gls{ssps} introduces very strong assumptions at the gNB, which makes it a suitable approach for benchmarking the performance of more practical schemes.
\smallskip

\subsubsection{\Acrfull{asps}}
\label{sub:semi-persistent gNB with learning}

To consider a more realistic scenario, we assume that the \gls{gnb} does not know a priori $\tau_{\rm on}$ and $n_{\rm on}$.
Hence, the \gls{gnb} implements an adaptive heuristic algorithm to estimate these missing parameters, as described in Algorithm~\ref{alg:algorithm1}. 
In particular, the \gls{gnb} initially estimates $\hat\tau_{\rm on}$ and $\hat{n}_{\rm on}$ based on the values of $T_{\rm IP}$ and $n_{\rm lines}$ (known).
Then, the value of $\hat n_{\rm on}$ is updated based on the ID of UEs sending scheduling requests in consecutive PUCCH opportunities, while the value of $\hat\tau_{\rm on}$ is updated based on the outcomes of previous resource allocations, that is based on the number of unused SUs in previous PUCCH opportunities.

Preliminary simulations show that a
training phase of at least $3T_{\rm IP}$ is required to accurately learn $\hat{\tau}_{\rm on}$, 
thus to perform error-free scheduling, which has implications in terms of latency, as we will show in Sec.~\ref{sec:numerical_results}.
Besides, during the training phase, some data blocks may be buffered, accumulating very long delays. For this reason, \glspl{ue} adopt a {dropping policy} that allows queued packets to be discarded if they are not transmitted within the current activation period (\gls{gi} included), thus promoting lower latency, at the cost of reduced reliability.

\begin{algorithm}[t!]
\caption{\Acrlong{asps} (Sec.~\ref{sub:semi-persistent gNB with learning})}
\label{alg:algorithm1}

\begin{algorithmic}[1]
\footnotesize
    \State \textit{System Initialization}
    \begin{itemize}
        \item $n_{\rm e}$: estimation cycle
        \item $\hat{\tau}_{\rm on}$: estimated activation period
        \item $\hat{n}_{\rm on}$: estimated number of activations
        \item $su_{\rm notx}$: list of \glspl{su} not used by \glspl{ue} for data tx
        \item $t_{\rm notx}[i]$: instant of time corresponding to $su_{\rm notx}[i]$
        \item $su_{\rm tx}$: list of \glspl{su} used by \glspl{ue} for data tx
        \item $t_{\rm tx}[i]$: instant of time corresponding to $su_{\rm tx}[i]$
    \end{itemize}
 
    \State \textit{Start} 
    \State $n_{\rm e} = 1$ $\rightarrow$ $\hat{\tau}_{\rm on} = T_{\rm IP}/{n_{\rm lines}}$ and $\hat{n}_{\rm on} ={n_{\rm lines}}$
    \State $n_{\rm e} = 2$ $\rightarrow$ \gls{gnb} recovers the activation period index ($k$) at $n_{\rm e}-1$ for \glspl{ue} that sent the  \gls{pucch} at $n_{\rm e}$ $\rightarrow {\hat n}_{\rm on} = n_{\rm lines} + k - 1$
    \State Based on $\hat{n}_{\rm on}$ and $T_{\rm IP}$ $\rightarrow \hat{\tau}_{\rm on}$ even ($\hat{\tau}_{\rm on_{\rm ev}}$) or odd ($\hat{\tau}_{\rm on_{\rm od}}$)
    \State \gls{gnb} keeps track of $su_{\rm notx}$
    \If{$\hat{\tau}_{\rm on_{\rm ev}} = \text{True}$}
        \State $\hat{\tau}_{\rm on} = t_{\rm {\rm notx}}[0] + (n_{\rm e}-1)\, (n_{\rm os})$
    \Else
        \State $\hat{\tau}_{\rm on} = t_{\rm {\rm notx}}[0]$ 
    \EndIf
    \State $n_{\rm e} = 3 \rightarrow$ \gls{gnb} keeps track of the new $su_{\rm notx}^{\rm (new)}$ and $su_{\rm tx}$
    \If{$su_{\rm notx}^{\rm (new)}[0] = su_{\rm notx}[0]$}
        \State $\hat{\tau}_{\rm on} = t_{\rm {notx}}[0] + (n_{\rm e}-1)\, (n_{\rm os})$ 
       
    \Else
        \For{$i = 1$ to lenght$(su_{\rm tx})$}
            \State $\Delta = su_{\rm tx}[{ i}] - su_{\rm tx}[ i-1]$
            \If{$\Delta > \hat{\tau}_{\rm on}$}
                \State $\hat{\tau}_{\rm on}^{\rm (new)} = \frac{\hat{\tau}_{\rm on}+t_{\rm tx}[i-1]}{2}$
                \State $\hat{\tau}_{\rm on} = \hat{\tau}_{\rm on}^{\rm (new)} + (n_{\rm e}-1)\, (n_{\rm os})$ 
            \EndIf
        \EndFor
    \EndIf

\end{algorithmic}
\end{algorithm}

\section{Performance Evaluation} 
\label{sec:numerical_results}
In Sec.~\ref{sub:numerical} we compare the \gls{e2e} latency, defined in Sec.~\ref{sec:e2e_latency}, of different \gls{sps} schemes, based on the simulation parameters introduced in Sec.~\ref{sub:params}.

\subsection{End-to-End Latency Analysis}
\label{sec:e2e_latency}
In this work 
we evaluate the \gls{e2e} latency experienced in the \gls{iiot} scenario referenced in Sec.~\ref{sec:e2e_system_model}.
The \gls{e2e} latency for a single data burst (\gls{sdu} at the application layer) is computed as the time between the generation of the data burst at the \gls{ue} and its reception at the \gls{cm}.
Formally, the \gls{e2e} latency $L$, is given by:\footnote{For a more detailed explanation of the terms in Eq.~\eqref{eq:e2e_latency_expression}, we refer the interested readers to the analysis in~\cite{eucnc}.} 
\begin{equation}
\label{eq:e2e_latency_expression}
    L = T_{\rm P} + T_{\rm RAN} + T_{\rm TX} + \tau_{\rm P} + T_{\rm FH} + \tau_{\rm FH} + T_{\rm gNB} + T_{\rm CN},
\end{equation}
where $T_{\rm RAN}$ is the time between the generation of the data block (\gls{pdu} at the physical layer) and its transmission over the channel, which depends on the scheduling algorithm, and on the system parameters. On the contrary, the other terms have constant values.
Finally, we introduce the average \gls{e2e} latency $\bar{L}$, averaged over the data blocks generated by the \glspl{ue} within the simulation time $T_S$, and over the number of \glspl{ue}.

\subsection{Simulation Parameters}
\label{sub:params}

\def\arraystretch{1.3}
\begin{table}[t!]
\centering
\footnotesize
\caption{Simulation parameters.}
\label{tab:system_parameter_settings}
\begin{tabular}{l|l|l}
\hline
\textbf{Parameter} & \textbf{Description} & \textbf{Value}  \\\hline

$f_{\rm c}$ & Carrier frequency & 3.5 GHz\\ 
$B$ & Overall system bandwidth & \{60,120\} MHz \\
$\Delta f$ & Subcarrier spacing & 60 kHz\\
$SNR_{\rm th}$ & SNR threshold & $-5$ dB  \\   
$T$ & Noise temperature & 290 K   \\
$T_{\rm P}$ & Delay to create data block &  7 \gls{ofdm} symbols\\
$T_{\rm TX}$ & Data block transmission time & 4 \gls{ofdm} symbols\\
$\tau_{\rm P}$ & Propagation time & 0\\
$T_{{\rm FH}}$ & {Fronthaul} delay & 0.05 ms \\
$\tau_{\rm FH}$ & FH-to-\gls{gnb} propagation time & 0\\
$T_{{\rm gNB}}$ & \gls{gnb} delay & 7 \gls{ofdm} symbols \\
$T_{\rm CN}$ & \gls{5gc} delay & 0.1 ms \\   
$T_{\rm S}$ & Simulation time & 10 s
\\  
$G_{\rm UE}=G_{\rm gNB}$ & Antenna gain & 0 dB \\   
$P_{\rm TX, UL}$ & UE (UL) transmit power & 23 dBm \\   
$P_{\rm TX, DL}$& gNB (DL) transmit power & 30 dBm \\  
$D$ & Inter-machine distance & \{5, 10\} m \\   
$S$ & Side of the machines & \{2, 3\} m \\  
$M_{\rm min}$ & Min. number of machines & 16 \\ 
$l$ & Length of the factory floor & \{20, 50\} m~\cite{5gaciaarchitecture}
\\   
$w$ & Width of the factory floor & \{20, 10\} m\\  
$h$ & Height of the factory floor & \{4, 10\} m \\  
$H$ & \gls{5g} protocol stack header & 72 bytes~\cite{etsi_tr_137_901_5}
\\  
$B_{\rm s}$ & Bucket size & 40\% of data burst 
\\
$p$ & User activation probability & 1\\\hline
\end{tabular}
\vspace{-3ex}
\end{table}

The simulation parameters are in Table~\ref{tab:system_parameter_settings}. We consider:
\begin{itemize}
    \item Two use cases, defined by the \gls{5gacia}: (1) augmented reality and (2) remote access and maintenance. According to the requirements in~\cite{5gaciarequirement}, and given their factory layouts, in (1) $n_{\rm lines} = 4$, with four machines per line, while in (2) $n_{\rm lines} = 2$, with eight machines per line.
    
    \item A subcarrier spacing of $\Delta_f=60$ kHz which, as observed in~\cite{eucnc}, offers lower latency compared to $\Delta_f=30$~kHz.

    \item Two values of the bandwidth, i.e., $B =$ 60 and 120 MHz, which lead to $n_{\rm RB} =$ 84 and 167 \glspl{rb}, respectively.
    \end{itemize}

    We compare the performance of 5G NR SPS presented in Sec.~\ref{sec:overview} and deployed in~\cite{eucnc} (referred to as \gls{bsps} in the rest of the paper) with our extensions \gls{ssps} and \gls{asps} for correlated traffic presented in Sec.~\ref{sec:semi-persistent_scheduling}.
Moreover, we analyze the performance of correlated traffic according to the model in Sec.~\ref{sub:correlated_traffic}, for both periodic and aperiodic data, as a function of the per-user offered traffic $G$, i.e., the ratio between the data block size and the traffic periodicity. 

\subsection{Numerical Results}
\label{sub:numerical}



\textbf{Impact of the SPS implementation.}
Tab.~\ref{tab:OSPS vs SSPS vs ASPS} compares the average \gls{e2e} latency for \gls{bsps}, \gls{ssps} and \gls{asps}, vs. the number of \glspl{ue}~$N$. We consider the augmented
reality use case, $\tau_{\rm on}=$ 8 ms, $B=$ 60 MHz, and $n_{\rm on}= 5$ (i.e., $T_{\rm IP}=$ 41.25 ms according to Eq.~\eqref{eq:t_ip}). 
Active \glspl{ue} generate periodic traffic with periodicity $\tau=2$ ms. 
We see that, for \gls{bsps}, $\overline{L}$ may grow indefinitely.
This is due to the fact that the \gls{us} can assign resources only to those \glspl{ue} making scheduling requests via the \gls{pucch}, while the others keep their data blocks in the queue, thus accumulating delays. 
On the other hand, \gls{ssps} and \gls{asps} are designed to predict the spatio-temporal correlations of the traffic even without explicit \gls{pucch} requests, and make faster scheduling decisions. While $\overline{L} \leq 1$ ms only for \gls{ssps} with $N<100$, and under the assumption that the gNB has complete information about the traffic statistics of machines and UEs, \gls{asps} can still provide very low latency (up to $-97\%$ compared to \gls{bsps}), even though during the training phase the \gls{us} may assign resources inaccurately, which implies that some data may remain in the queue and accumulate delays. 
\smallskip
\begin{table}[!t]
\centering
\renewcommand{\arraystretch}{1.3}
\caption{\gls{e2e} latency vs. N, for different SPS implementations.}
\label{tab:OSPS vs SSPS vs ASPS}
\begin{tabular}{c|c|c|c}
\hline
{$N$} & {$\overline{L}_{\rm BSPS}$ [ms]} & {$\overline{L}_{\rm SSPS}$ [ms]} & {$\overline{L}_{\rm ASPS}$ [ms]}  \\\hline
60 & 97.2 & 0.798 & 4.417\\ 
70 & 107.429 & 0.846 & 4.782 \\
80 & 118.774 & 0.901 & 4.865 \\
90 & 131.056 & 0.967 & 4.998\\   
100 & 147.682 & 1.022 & 5.089\\
\hline
\end{tabular}

\end{table}
\begin{figure}[t!]
\begin{center}
    \includegraphics[width = 0.8\columnwidth]{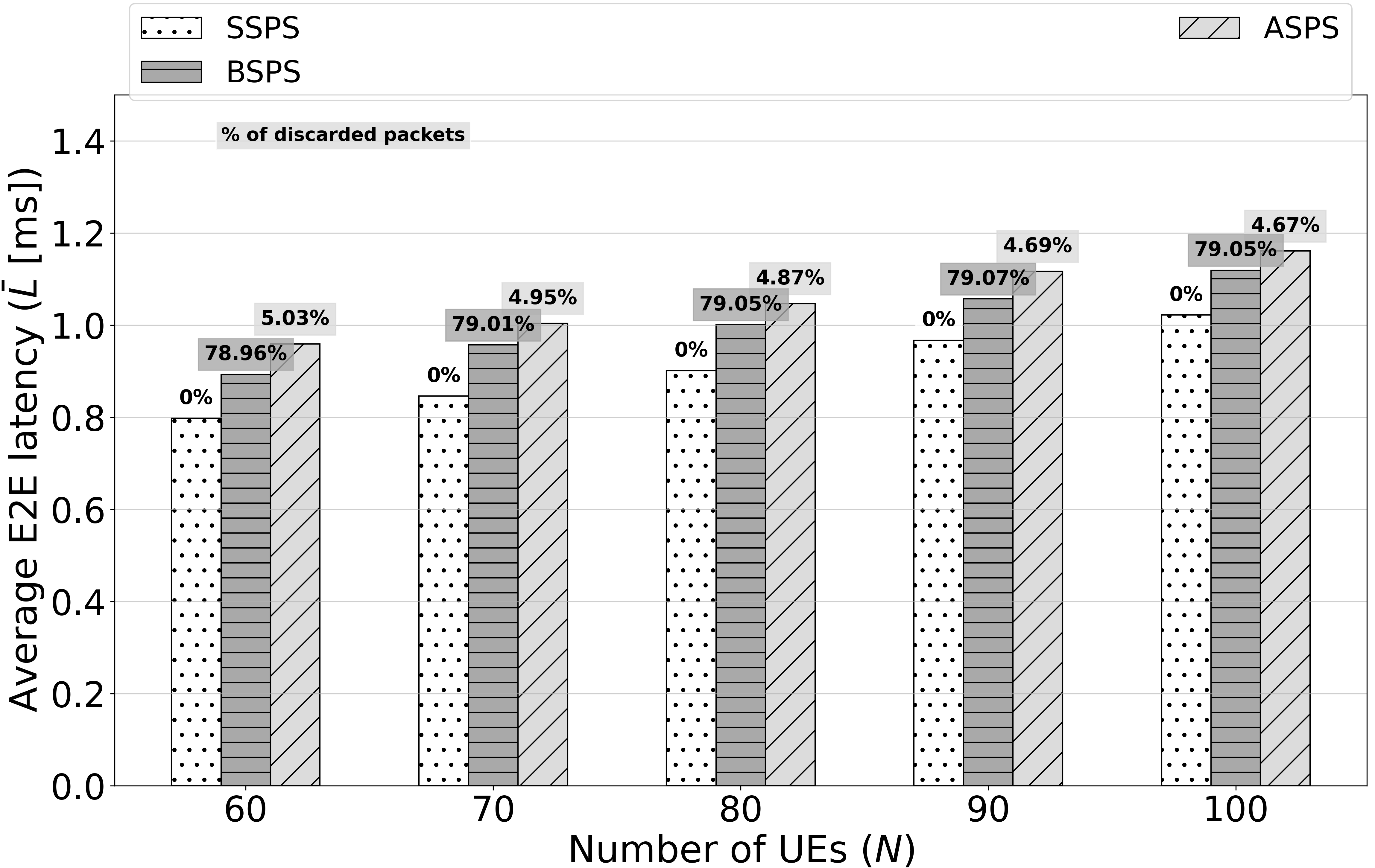}
    \caption{Average E2E latency vs. $N$, for \gls{bsps}, \gls{ssps} and \gls{asps} with the dropping policy. We consider the augmented reality use case, $G =$ 2.75 Mbit/s, $\tau_{\rm on} =$ 8 ms, $n_{\rm on} =$ 5. Numbers on top of the bars are the packet loss ratios.}
    \label{fig:OPS_SSPS_ASPS}
\end{center}
\vspace{-3ex}
\end{figure}

\textbf{Impact of the dropping policy.}
In Fig.~\ref{fig:OPS_SSPS_ASPS} we still evaluate the \gls{e2e} latency for BSPS, SSPS and ASPS, but now we introduce the dropping policy as described in Sec.~\ref{sub:semi-persistent gNB with learning}. 
While \gls{ssps} is designed to schedule resources with 100\% accuracy, unscheduled packets in \gls{bsps} and \gls{asps} will now be discarded if they are not transmitted within the UE's activation period. 
The dropping policy is the result of a compromise between latency and reliability:
for \gls{asps}, we can see that $\overline{L}$ is up to 75\% lower than the results in Tab.~\ref{tab:OSPS vs SSPS vs ASPS} (that do not include the dropping policy), and can satisfy the 1-ms latency requirement for \gls{urllc} as long as $N<70$, despite some packet loss. 
Notice that the latency for \gls{bsps} is not meaningful, as it comes with around 80\% packet loss, which makes the system less
congested: the (few) packets that will not be discarded because of the dropping policy are then transmitted with lower delay. As such, we can conclude that \gls{bsps} is totally unreliable for correlated traffic. 
\smallskip

\begin{figure}[t!]
\begin{center}
    \includegraphics[width = 0.8\columnwidth]{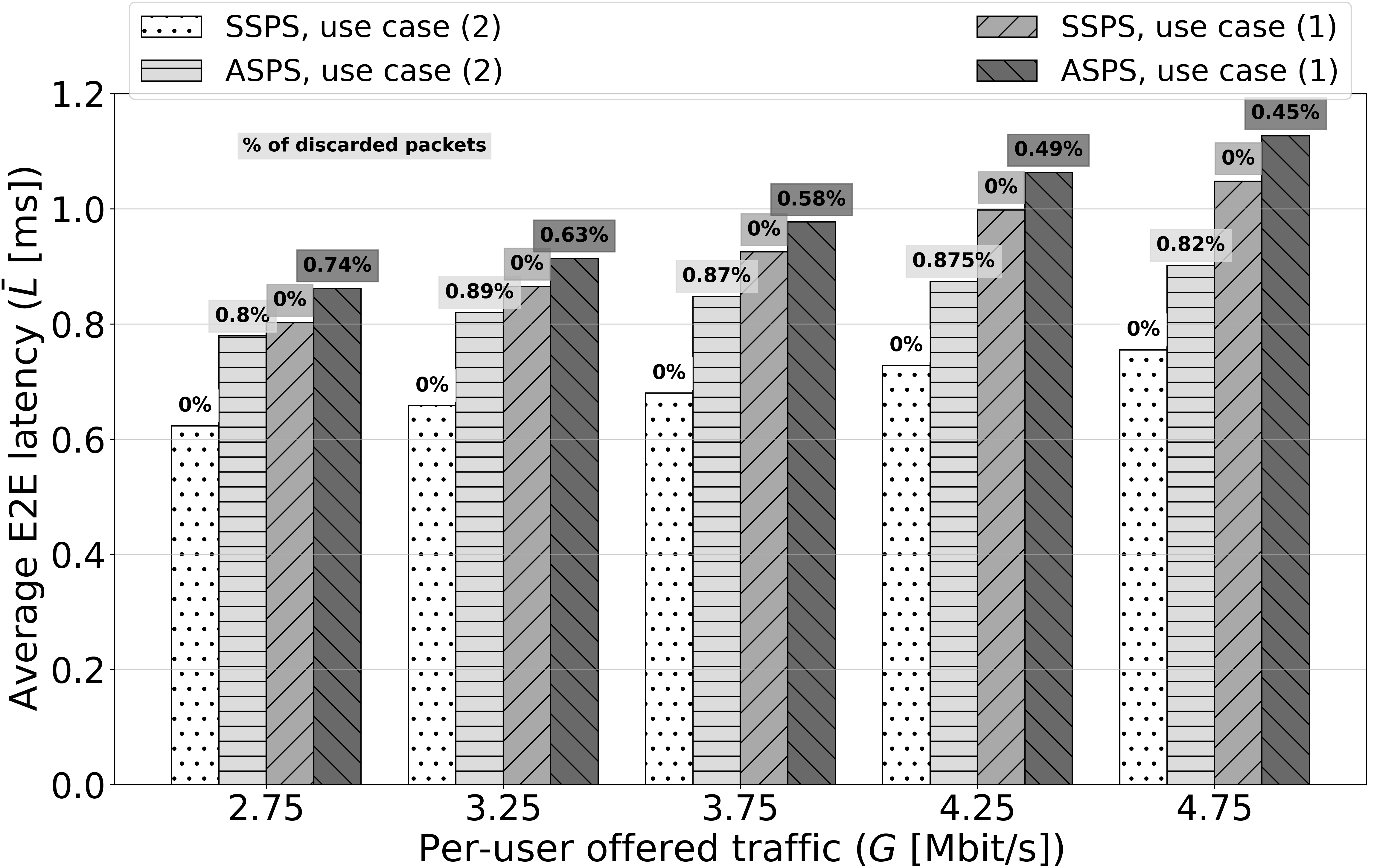}
    \caption{Average E2E latency vs. $G$ and the use case (where (1) is for augmented reality and (2) is for remote access and maintenance), for \gls{ssps} and \gls{asps} with the dropping policy. We set $N =$ 60, $\tau_{\rm on} =$ 8 ms. Numbers on top of the bars are the packet loss ratios.}
    \label{fig:SPSvsASPS_2usecases}
\end{center}
\vspace{-2ex}
\end{figure}

\textbf{Impact of the use case and $G$.} 
Fig.~\ref{fig:SPSvsASPS_2usecases} shows the average \gls{e2e} latency for \gls{ssps} and \gls{asps} with the dropping policy, as a function of $G$ and of the use case. 
According to Eq.~\eqref{eq:t_ip}, and given that $n_{\rm on}$ is equal to the number of machines per production line, we have that for use case (1) (augmented reality) $T_{\rm IP}=33$ ms, while for use case (2) (remote access and maintenance) $T_{\rm IP}=66$ ms.
As expected, the \gls{e2e}~latency increases with $G$ since the network is more congested. Also, the impact of the use case is remarkable: the number of machines (and, consequently, of \glspl{ue}) that can be active simultaneously is proportional to $n_{\rm lines}$, and is higher for use case (1), meaning that each \gls{ue} can be assigned fewer \glspl{rb} on average.
In these conditions, the 1-ms requirement for \gls{urllc} can always be satisfied for use case (2), while for (1) it should be $G\leq4.25$ Mbit/s if \gls{ssps} or $G\leq3.75$ Mbit/s if~\gls{asps}.
\smallskip

\begin{figure}[t!]
\begin{center}
    \includegraphics[width = 0.8\columnwidth]{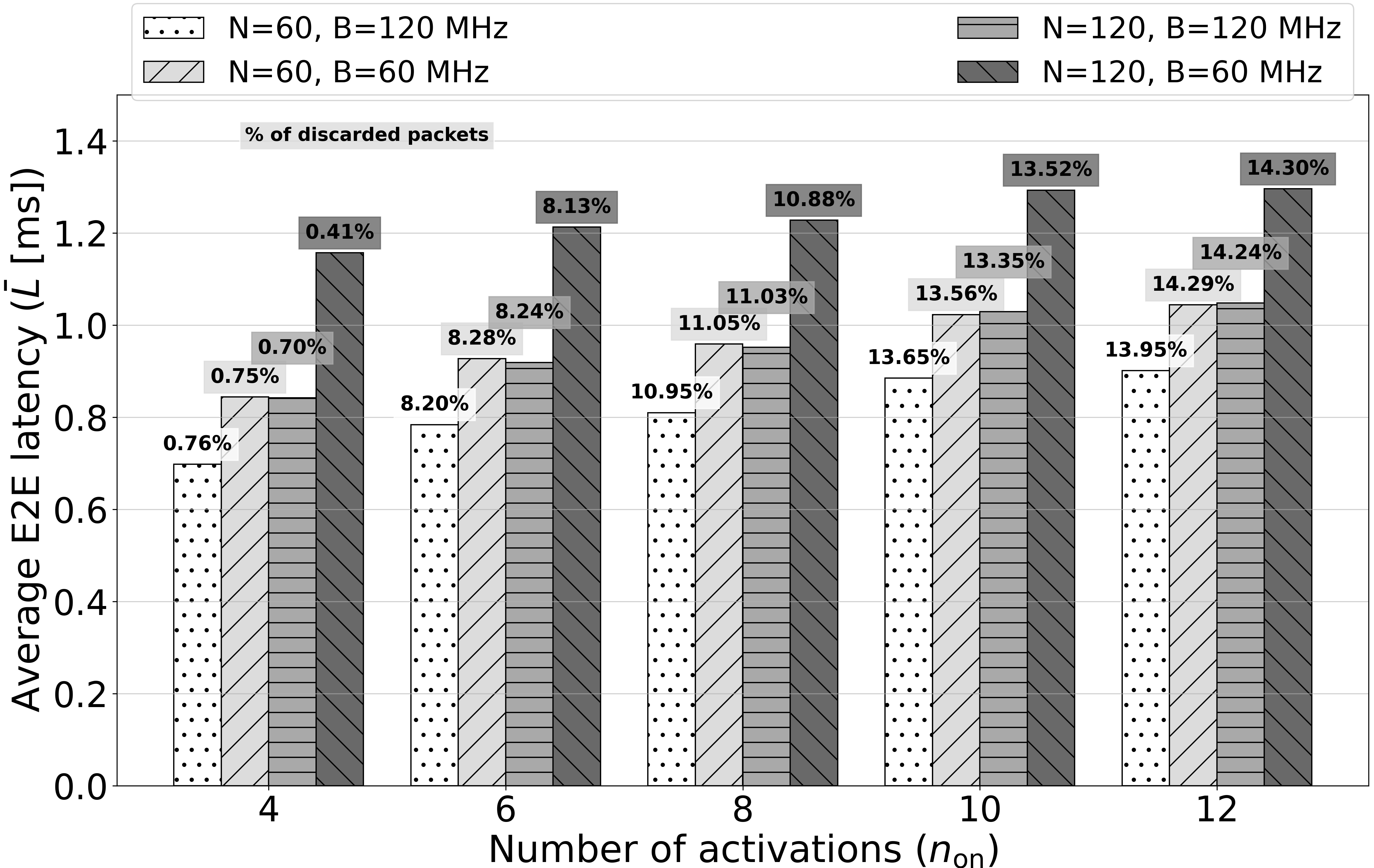}
    \caption{Average E2E latency vs. $n_{\rm on}$, $N$ and $B$ for \gls{asps} with the dropping policy. We consider the augmented reality use case, $\tau_{\rm on} =$ 8 ms, $G =$ 2.75 Mbit/s. Numbers on top of the bars are the packet loss ratios.}
    \label{fig:periodic_learning}
\end{center}
\vspace{-3ex}
\end{figure}

\textbf{Impact of the number of activations and $B$.} Fig.~\ref{fig:periodic_learning} shows the average \gls{e2e} latency for \gls{asps} with the dropping policy, as a function of $n_{\rm on}$, $B$ and $N$, for the augmented reality use case and $G =$ 2.75 Mbit/s.
We observe that $\overline{L}$ increases as $B$ decreases, proportionally with $N$. 
When $N=120$, the 1-ms \gls{urllc} latency requirement can never be achieved for $B =60$ MHz since $n_{\rm RB}$ is lower than $N$, or only when $n_{\rm on}\leq8$ for $B =120$ MHz.
When $N=60$, $\overline{L}$ is similar in both configurations, meaning that the impact of the bandwidth is negligible if the ratio between $N$ and $B$ is constant.

Moreover, $\overline{L}$ grows with the number of activations: at the beginning of the training, ASPS may underestimate $n_{\rm on}$, which implies that many data blocks would not be scheduled or, equivalently, some \glspl{ue} will be assigned fewer \glspl{rb} than expected, which would increase the transmission delay. 
Besides, the probability of underestimating ${n}_{\rm on}$ increases as the actual $n_{\rm on}$ increases, which explains the increasing trend of the packet loss in Fig.~\ref{fig:periodic_learning}.
\smallskip

\begin{figure}[t!]
\begin{center}
    \includegraphics[width = 0.8\columnwidth]{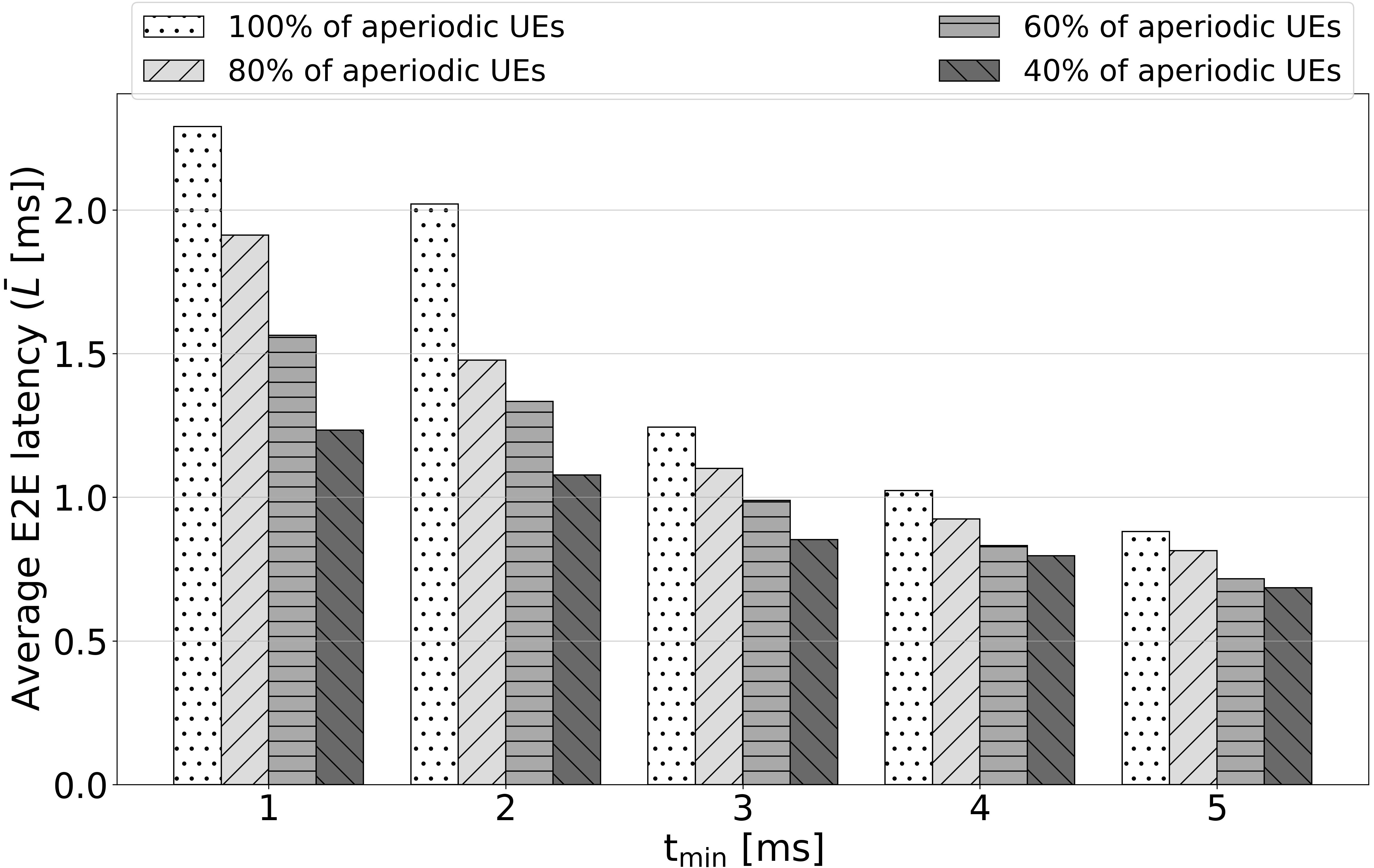}
    \caption{Average E2E latency vs. $t_{\rm min}$ and the percentage of aperiodic \glspl{ue} for ASPS with the dropping policy. We set $N =$ 60,  $\tau_{\rm on} =$ 8 ms, $B =$ 60 MHz, $G =$ 2.75 Mbit/s, $t_{\rm max} =$ 6 ms. Numbers on top of the bars are the packet loss ratios.}
    \label{fig:aperiodic_all_info}
\end{center}
\vspace{-3ex}
\end{figure}

\textbf{Impact of aperiodic traffic.} 
Since the \gls{gnb} has complete system information in \gls{ssps}, we explored the impact of aperiodic traffic on the \gls{e2e} latency in the best case.
The results are illustrated in Fig.~\ref{fig:aperiodic_all_info} as a function of $t_{\rm min}$. 
We assume that $n_{\rm on}$ is equal to the number of production lines, and that \{100\%, 80\%, 60\%, 40\%\} of the \glspl{ue} generate aperiodic traffic with $t_{\rm max}=6$~ms. 
We observe that $\overline{L}$ increases when $t_{\rm min}$ decreases because the traffic is more intense and the system is more congested.
However, the 1-ms requirement for \gls{urllc} cannot be satisfied in most cases when introducing totally unpredictable aperiodic traffic: the requirement is met with 100\% of aperiodic \glspl{ue} only when $t_{\rm min} = 5 $ ms. 
This suggests that \gls{sps} (also in the best case of \gls{ssps}) is not always compatible with unpredictable traffic, 
even considering some degrees of correlation in the activation of machines and UEs, which motivates further studies towards more sophisticated (e.g., grant-free/distributed) scheduling methods~\cite{pase2022distributed}.

\section{Conclusions}
In this work we presented two new custom designs of the SPS (\gls{ssps} and \gls{asps}, with the former that serves as a benchmark for the latter) that allocate resources in a \gls{5gacia} \gls{snpn} architecture, considering IIoT-specific spatio-temporal traffic correlations.
For both, we assessed the \gls{e2e} latency vs. a baseline 5G NR scheduling implementation (BSPS), as a function of the bandwidth, the use case, the per-user offered traffic, and the number of \glspl{ue}. 
Simulation results show that \gls{asps} significantly outperforms BSPS, and can satisfy the 1-ms URLLC requirement in many configurations. 
However, for aperiodic traffic, no scheme can strictly support URLLC. As such, as part of our future work, we will explore the design of more advanced (learning-based) scheduling techniques to speed up the training of \gls{asps} and achieve faster resource allocation, with considerations related to energy~consumption.
\label{sec:conclusions}

\section{Acknowledgment}
This work has been carried out in the framework of the CNIT/WiLab-Huawei Joint Innovation Center.

It was also partially supported by the European Union under the Italian National Recovery and Resilience Plan (NRRP) of NextGenerationEU, 
partnership on “Telecommunications of the Future” (PE0000001 - program “RESTART”).

\bibliographystyle{IEEEtran}
\bibliography{bibl.bib}

\end{document}

%% file: acronyms.tex
\newacronym{3gpp}{3GPP}{3rd Generation Partnership Project}
\newacronym{adc}{ADC}{Analog to Digital Converter}
\newacronym{5g}{5G}{5th generation}
\newacronym{aimd}{AIMD}{Additive Increase Multiplicative Decrease}
\newacronym{am}{AM}{Acknowledged Mode}
\newacronym{amc}{AMC}{Adaptive Modulation and Coding}
\newacronym{aqm}{AQM}{Active Queue Management}
\newacronym{awgn}{AWGN}{Additive White Gaussian Noise}
\newacronym{balia}{BALIA}{Balanced Link Adaptation}
\newacronym{bdp}{BDP}{Bandwidth-Delay Product}
\newacronym{qos}{QoS}{Quality of Service}
\newacronym{pqos}{PQoS}{predictive quality of service}
\newacronym{bf}{BF}{Beamforming}
\newacronym{cc}{CC}{Congestion Control}
\newacronym{cdf}{CDF}{Cumulative Distribution Function}
\newacronym{cn}{CN}{Core Network}
\newacronym{cqi}{CQI}{Channel Quality Information}
\newacronym{cp}{CP}{Control Plane}
\newacronym{csirs}{CSI-RS}{Channel State Information - Reference Signal}
\newacronym{dc}{DC}{Dual Connectivity}
\newacronym{dce}{DCE}{Direct Code Execution}
\newacronym{dci}{DCI}{Downlink Control Information}
\newacronym{dl}{DL}{Downlink}
\newacronym{dmr}{DMR}{Deadline Miss Ratio}
\newacronym{dmrs}{DMRS}{DeModulation Reference Signal}
\newacronym{e2e}{E2E}{end-to-end}
\newacronym{ecn}{ECN}{Explicit Congestion Notification}
\newacronym{edf}{EDF}{Earliest Deadline First}
\newacronym{enb}{eNB}{evolved Node Base}
\newacronym{epc}{EPC}{Evolved Packet Core}
\newacronym{es}{ES}{Edge Server}
\newacronym{fdma}{FDMA}{Frequency Division Multiple Access}
\newacronym{fdd}{FDD}{Frequency Division Duplexing}
\newacronym[firstplural=Radio Access Technologies (RATs)]{rat}{RAT}{Radio Access Technology}
\newacronym{fs}{FS}{Fast Switching}
\newacronym{ftp}{FTP}{File Transfer Protocol}
\newacronym{gnb}{gNB}{Next Generation NodeB}
\newacronym{harq}{HARQ}{Hybrid Automatic Repeat reQuest}
\newacronym{hetnet}{HetNet}{Heterogeneous Network}
\newacronym{hh}{HH}{Hard Handover}
\newacronym{hol}{HOL}{Head-of-Line}
\newacronym{ia}{IA}{Initial Access}
\newacronym{ieee}{IEEE}{Institute of Electrical and Electronics Engineers}
\newacronym{imt}{IMT}{International Mobile Telecommunication}
\newacronym{iot}{IoT}{Internet of Things}
\newacronym{ldpc}{LDPC}{Low-Density Parity Check}
\newacronym{los}{LOS}{Line-of-Sight}
\newacronym{lte}{LTE}{Long Term Evolution}
\newacronym{m2m}{M2M}{Machine to Machine}
\newacronym{ml}{ML}{machine learning}
\newacronym{mac}{MAC}{Medium Access Control}
\newacronym{mc}{MC}{Multi-Connectivity}
\newacronym{mcs}{MCS}{Modulation and Coding Scheme}
\newacronym{mec}{MEC}{Mobile Edge Cloud}
\newacronym{mi}{MI}{Mutual Information}
\newacronym{mimo}{MIMO}{Multiple Input, Multiple Output}
\newacronym{mmwave}{mmWave}{millimeter wave}
\newacronym{mptcp}{MPTCP}{Multipath TCP}
\newacronym{mr}{MR}{Maximum Rate}
\newacronym{mss}{MSS}{Maximum Segment Size}
\newacronym{mtd}{MTD}{Machine-Type Device}
\newacronym{mtu}{MTU}{Maximum Transmission Unit}
\newacronym{nfv}{NFV}{Network Function Virtualization}
\newacronym{nlos}{NLOS}{Non-Line-of-Sight}
\newacronym{nlosv}{NLOSv}{Vehicle Non-Line-of-Sight}
\newacronym{nr}{NR}{New Radio}
\newacronym{ofdm}{OFDM}{Orthogonal Frequency Division Multiplexing}
\newacronym{pdcch}{PDCCH}{Physical Downlink Control Channel}
\newacronym{sps}{SPS}{semi-persistent scheduler}
\newacronym{pdcp}{PDCP}{Packet Data Convergence Protocol}
\newacronym{pdsch}{PDSCH}{Physical Downlink Shared Channel}
\newacronym{pdu}{PDU}{Packet Data Unit}
\newacronym{pf}{PF}{Proportional Fair}
\newacronym{ff}{FF}{Fairness First}
\newacronym{gi}{GI}{guard interval}

\newacronym{pgw}{PGW}{Packet Gateway}
\newacronym{phy}{PHY}{Physical}
\newacronym{pbch}{PBCH}{Physical Broadcast Channel}
\newacronym[plural=\gls{mme}s,firstplural=Mobility Management Entities (MMEs)]{mme}{MME}{Mobility Management Entity}
\newacronym{prb}{PRB}{Physical Resource Block}
\newacronym{pss}{PSS}{Primary Synchronization Signal}
\newacronym{pscch}{PSCCH}{Physical Sidelink Control Channel}
\newacronym{pucch}{PUCCH}{Physical Uplink Control Channel}
\newacronym{pusch}{PUSCH}{Physical Uplink Shared Channel}
\newacronym{rach}{RACH}{Random Access Channel}
\newacronym{ran}{RAN}{Radio Access Network}
\newacronym{red}{RED}{Random Early Detection}
\newacronym{rf}{RF}{radio frequency}
\newacronym{rlc}{RLC}{Radio Link Control}
\newacronym{rlf}{RLF}{Radio Link Failure}
\newacronym{rrc}{RRC}{Radio Resource Control}
\newacronym{rrm}{RRM}{Radio Resource Management}
\newacronym{rr}{RR}{Round Robin}
\newacronym{rs}{RS}{Remote Server}
\newacronym{rsrp}{RSRP}{Reference Signal Received Power}
\newacronym{rss}{RSS}{Received Signal Strength}
\newacronym{rtt}{RTT}{Round Trip Time}
\newacronym{rw}{RW}{Receive Window}
\newacronym{rx}{RX}{Receiver}
\newacronym{sa}{SA}{standalone}
\newacronym{sack}{SACK}{Selective Acknowledgment}
\newacronym{sap}{SAP}{Service Access Point}
\newacronym{sc}{SC}{Single Carrier}
\newacronym{sch}{SCH}{Secondary Cell Handover}
\newacronym{scoot}{SCOOT}{Split Cycle Offset Optimization Technique}
\newacronym{sdma}{SDMA}{Spatial Division Multiple Access}
\newacronym{sinr}{SINR}{Signal to Interference plus Noise Ratio}
\newacronym{sl}{SL}{Sidelink}
\newacronym{sm}{SM}{Saturation Mode}
\newacronym{snr}{SNR}{Signal-to-Noise-Ratio}
\newacronym{son}{SON}{Self-Organizing Network}
\newacronym{ss}{SS}{Synchronization Signal}
\newacronym{srs}{SRS}{Sounding Reference Signal}
\newacronym{sss}{SSS}{Secondary Synchronization Signal}
\newacronym{tb}{TB}{Transport Block}
\newacronym{tcp}{TCP}{Transmission Control Protocol}
\newacronym{tdd}{TDD}{Time Division Duplexing}
\newacronym{tdma}{TDMA}{Time Division Multiple Access}
\newacronym{tfl}{TfL}{Transport for London}
\newacronym{tm}{TM}{Transparent Mode}
\newacronym{trp}{TRP}{Transmitter Receiver Pair}
\newacronym{tti}{TTI}{Transmission Time Interval}
\newacronym{ttt}{TTT}{Time-to-Trigger}
\newacronym{tx}{TX}{Transmitter}
\newacronym{ue}{UE}{User Equipment}
\newacronym{ul}{UL}{uplink}
\newacronym{uml}{UML}{Unified Modeling Language}
\newacronym{um}{UM}{Unacknowledged Mode}
\newacronym{utc}{UTC}{Urban Traffic Control}
\newacronym{vm}{VM}{Virtual Machine}
\newacronym{rsrq}{RSRQ}{Reference Signal Received Quality}
\newacronym{rssi}{RSSI}{Received Signal Strength Indicator}
\newacronym{crs}{CRS}{Cell Reference Signal}
\newacronym{nsa}{NSA}{Non Stand Alone}
\newacronym{mrdc}{MR-DC}{Multi \gls{rat} \gls{dc}}
\newacronym{endc}{EN-DC}{E-UTRAN-\gls{nr} \gls{dc}}
\newacronym{5gc}{5GC}{5G Core}
\newacronym{si}{SI}{Study Item}
\newacronym{iab}{IAB}{Integrated Access and Backhaul}
\newacronym{wf}{WF}{Wired-first}
\newacronym{hqf}{HQF}{Highest-quality-first}
\newacronym{pa}{PA}{Position-aware}
\newacronym{mlr}{MLR}{Maximum-local-rate}
\newacronym{wbf}{WBF}{Wired Bias Function}
\newacronym{mib}{MIB}{Master Information Block}
\newacronym{sib}{SIB}{Secondary Information Block}
\newacronym{rnti}{RNTI}{Radio Network Temporary Identifier}
\newacronym{dft}{DFT}{Discrete Fourier Transform}
\newacronym{kpi}{KPI}{Key Performance Indicator}
\newacronym{ppp}{PPP}{Poisson Point Process}
\newacronym{v2v}{V2V}{Vehicle-to-Vehicle}
\newacronym{wave}{WAVE}{Wireless Access in Vehicular Environments}
\newacronym{udp}{UDP}{User Datagram Protocol}
\newacronym{upa}{UPA}{Uniform Planar Array}
\newacronym{fec}{FEC}{Forward Error Correction}
\newacronym{v2x}{V2X}{Vehicle-To-Everything}
\newacronym{psfch}{PSFCH}{Physical Sidelink Feedback Channel}
\newacronym{pssch}{PSSCH}{Physical Sidelink Shared Channel}
\newacronym{csma}{CSMA}{Carrier Sense Multiple Access}
\newacronym{v2n}{V2N}{Vehicle-to-Network}
\newacronym{wlan}{WLAN}{Wireless Local Area Network}
\newacronym{cav}{CAV}{Connected and Autonomous Vehicle}
\newacronym{v2i}{V2I}{Vehicle-to-Infrastructure}
\newacronym{d2d}{D2D}{Device-to-Device}
\newacronym{c-its}{C-ITS}{Connected Intelligent Transportation System}
\newacronym{fr2}{FR2}{Frequency Range 2}
\newacronym{fr1}{FR1}{Frequency Range 1}
\newacronym{bs}{BS}{Base Station}
\newacronym{sdu}{SDU}{Service Data Unit}
\newacronym{csi}{CSI}{Channel State Information}
\newacronym{scs}{SCS}{Subcarrier Spacing}
\newacronym{sumo}{SUMO}{Simulation of Urban MObility}
\newacronym{prr}{PRR}{Packet Reception Ratio}
\newacronym{edca}{EDCA}{Enhanced Distribution Channel Access}
\newacronym{sdap}{SDAP}{Service Data Adaptation Protocol}
\newacronym{iiot}{IIoT}{Industrial Internet of Things}
\newacronym{agv}{AGV}{Automated Guided Vehicles}
\newacronym{cm}{C/M}{Controller/Master}
\newacronym{soa}{SoA}{State-of-the-Art}
\newacronym{snpn}{SNPN}{Standalone Non-Public Network}
\newacronym{pninpn}{PNI-NPN}{Public Network Interface Non-Public Network}
\newacronym{urllc}{URLLC}{Ultra-Reliable Low-Latency Communications}
\newacronym{ai}{AI}{Artificial Intelligence}
\newacronym{mab}{MAB}{Multi-Armed Bandit}
\newacronym{su}{SU}{Scheduling Unit}
\newacronym{ra}{RA}{Random Agent}
\newacronym{na}{NA}{Neural Agent}
\newacronym{ucba}{UCB-A}{UCB Agent}
\newacronym{tsa}{TS-A}{TS Agent}
\newacronym{ucb}{UCB}{Upper Confidence Bound}
\newacronym{ts}{TS}{Thompson Sampling}
\newacronym{nlts}{NLTS}{Neural Linear Thompson Sampling}
\newacronym{inf}{InF}{Indoor Factory}
\newacronym{infsl}{InF-SL}{Indoor Factory - Sparse Clutter, Low BS}
\newacronym{infdl}{InF-DL}{Indoor Factory - Dense Clutter, Low BS}
\newacronym{infsh}{InF-SH}{Indoor Factory - Sparse Clutter, High BS}
\newacronym{infdh}{InF-DH}{Indoor Factory - Dense Clutter, High BS}
\newacronym{us}{US}{Uplink Scheduler}
\newacronym{nn}{NN}{Neural Network}
\newacronym{dnn}{DNN}{Deep Neural Network}
\newacronym{das}{DAS}{Distributed Antenna System}
\newacronym{rb}{RB}{Resource Block}
\newacronym{rl}{RL}{Reinforcement Learning}
\newacronym{embb}{eMBB}{enhanced Mobile BroadBand}
\newacronym{5gacia}{5G-ACIA}{5G Alliance for Connected Industries and Automation}
\newacronym{cnn}{CNN}{Convolutional Neural Network}
\newacronym{aipm}{AIPM}{Artificial Intelligence Program Manager}
\newacronym{fci}{FCI}{Feedback Control Information}
\newacronym{prach}{PRACH}{Physical Random Access Channel}
\newacronym{coreset}{CORESET}{Control Resource Set}
\newacronym{cmab}{CMAB}{Contextual Multi-Armed Bandit}
\newacronym{drl}{DRL}{Deep Reinforcement Learning}
\newacronym{uav}{UAV}{Unmanned Aerial Vehicle}
\newacronym{ssps}{SSPS}{smart semi-persistent scheduler}
\newacronym{asps}{ASPS}{adaptive semi-persistent scheduler}
\newacronym{osps}{OSPS}{original semi-persistent scheduler}
\newacronym{bsps}{BSPS}{baseline semi-persistent scheduler}
\newacronym{fd}{FD}{frequency domain}
\newacronym{td}{TD}{time domain}